\documentclass[envcountsame,runningheads,notitlepage]{llncs}

\usepackage{cite}
\usepackage{amsmath,amssymb,amsfonts}
\usepackage{algorithmic}
\usepackage{graphicx}
\usepackage{textcomp}
\usepackage{xcolor}

\usepackage{array}
\usepackage{multirow}
\usepackage{booktabs}

\def\BibTeX{{\rm B\kern-.05em{\sc i\kern-.025em b}\kern-.08em
    T\kern-.1667em\lower.7ex\hbox{E}\kern-.125emX}}

\begin{document}

\title{Cross-Platform Benchmarking of the FHE Libraries: Novel Insights into SEAL and OpenFHE}

\authorrunning{Faneela, Ahmad, Ghaleb, Jan, and Buchanan}
\titlerunning{Cross-Platform Benchmarking of the FHE}

\author{%
Faneela\inst{1}  \and 
Jawad Ahmad\inst{2} \and
Baraq Ghaleb\inst{1} \and
Sana Ullah Jan\inst{1} \and
William J. Buchanan\inst{1}
}

\institute{Blockpass ID Lab, Edinburgh Napier University, UK\and Cybersecurity Center, Prince Mohammad Bin Fahd University, Alkhobar, Saudi Arabia}

\maketitle

\begin{abstract}
The rapid growth of cloud computing and data-driven applications has amplified privacy concerns, driven by the increasing demand to process sensitive data securely. Homomorphic encryption (HE) has become a vital solution for addressing these concerns by enabling computations on encrypted data without revealing its contents. This paper provides a comprehensive evaluation of two leading HE libraries, SEAL and OpenFHE, examining their performance, usability, and support for prominent HE schemes such as BGV and CKKS.   Our analysis highlights computational efficiency, memory usage, and scalability across Linux and Windows platforms, emphasizing their applicability in real-world scenarios. Results reveal that Linux outperforms Windows in computation efficiency, with OpenFHE emerging as the optimal choice across diverse cryptographic settings. This paper provides valuable insights for researchers and practitioners to advance privacy-preserving applications using FHE.
\end{abstract}

\begin{keywords}
Homomorphic encryption, HE libraries, cross-platform, resource utilization, privacy-preserving
\end{keywords}

\section{Introduction}
 Cloud computing has revolutionized modern computing by providing network access through cloud servers, which are remote storage systems that facilitate data processing and reduce the computational burden on local networks. While cloud servers offer significant advantages in terms of efficiency and scalability, their involvement also introduces critical security concerns, as sensitive data is entrusted to third-party entities. \cite{munjal2023systematic}. To address the security concerns associated with cloud computing, encryption techniques are employed to transform sensitive data into an unreadable form, ensuring its protection during transmission and storage.  While essential for securing data during transmission and storage, traditional encryption techniques  do  not not fully address the privacy concerns in cloud computing as they typically require data to be decrypted before any computation can occur, exposing sensitive information to potential vulnerabilities during processing. This limitation underscores the need for more advanced encryption methods, such as Homomorphic Encryption (HE), which enables computations to be performed directly on encrypted data without the need for decryption. HE resolves the privacy challenges by allowing data to remain confidential throughout its lifecycle, from storage to computation. Figure 1 illustrates the distinction between conventional cloud computing and HE-based computing, emphasizing the enhanced security and privacy protection provided by HE.
\begin{figure*}[ht]
    \centering
    \includegraphics[width=0.85\textwidth]{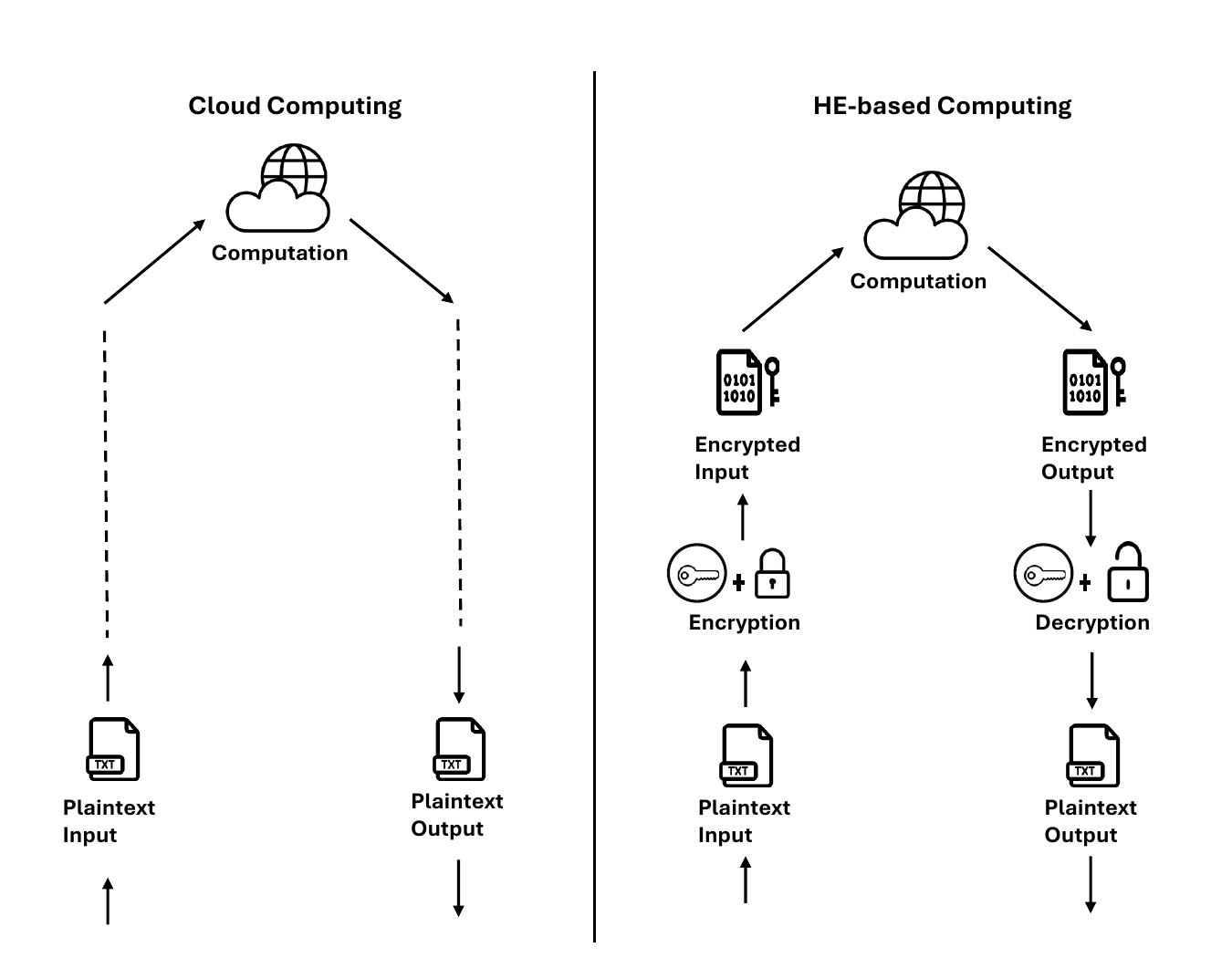} 
    \caption{Architectural Difference between Cloud and HE-based Computing}
    \label{fig:fig1}
\end{figure*}

Homomorphic Encryption (HE) has made significant progress since its inception, evolving from a concept that allowed only simple arithmetic operations on encrypted data to a powerful tool for secure computations on large-scale data. Initially, HE was limited to performing either addition or multiplication on encrypted data, restricting its practical applicability. However, the field underwent a major breakthrough with Gentry's introduction of Fully Homomorphic Encryption (FHE) in 2009, which enabled both addition and multiplication operations on encrypted data simultaneously. This advancement opened the door to more complex and scalable computations, making HE suitable for a broader range of real-world applications \cite{marcolla2022survey}. In his research, he addressed the limitation of initial HE by introducing both types of homomorphic operations on encrypted data. Despite its theoretical promise, FHE faced challenges in practical implementation due to its high computational complexity. In response, several specialized libraries, such as Microsoft SEAL and OpenFHE, were developed, each offering efficient implementations of different FHE schemes, including the BGV and CKKS schemes. These libraries have made it possible to perform precise integer computations (BGV) or approximate floating-point operations (CKKS), further expanding the versatility and feasibility of HE in privacy-preserving computing. As the adoption of Homomorphic Encryption (HE) grows, it becomes essential to thoroughly analyze and evaluate the available HE libraries. Indeed,  each library, such as Microsoft SEAL and OpenFHE, offers different trade-offs in terms of computational efficiency, ease of use, and support for various HE schemes. Therefore, a comprehensive evaluation is necessary to identify the strengths and limitations of these libraries in real-world scenarios.  This evaluation will help researchers, developers, and organizations make informed decisions about which library best meets their specific needs, whether it be for secure data processing in cloud environments or for implementing scalable privacy-preserving systems. Furthermore, analyzing these libraries will provide valuable insights into their optimization and potential areas for improvement, ensuring that HE can be effectively deployed in practical, large-scale applications.
\begin{enumerate}
    \item How do FHE libraries perform across different configuration settings?
    \item  How do varying HE parameters impact the performance of FHE libraries?
    \item How does the choice of platform (Linux vs. Windows) influence the performance of FHE libraries?
    \item How do different FHE schemes (e.g., BGV vs. CKKS) perform across various FHE libraries and OS platforms?
\end{enumerate}
Hence, our study introduces a novel benchmark for evaluating the performance of Microsoft SEAL and OpenFHE, focusing on the BGV and CKKS schemes. We specifically examine two key FHE operations—homomorphic addition and multiplication. Additionally, we conduct a cross-platform analysis to assess the performance of these libraries in various environments.

\section{Background}
Practical implementations of Homomorphic Encryption (HE) have faced several challenges, including mathematical complexity and increased resource demands. To address these limitations, various encryption schemes have been proposed, such as RSA, ElGamal, and Paillier. In 1977, Rivest-Shamir-Adleman (RSA) proposed a practical method of HE using multiplication functions on encrypted messages, $m_1$, and $m_2$, such as:
\begin{equation}
\text{Enc}(m_1) \cdot \text{Enc}(m_2) = \text{Enc}(m_1 \cdot m_2)
\end{equation}
Similarly, Paillier introduced an additive version of RSA’s scheme where the operation is:
\begin{equation}
\text{Enc}(m_1) + \text{Enc}(m_2) = \text{Enc}(m_1 + m_2)
\end{equation}

However, both RSA and Paillier schemes were limited to performing only one type of homomorphic operation—either addition or multiplication \cite{mohammed2022performance}, \cite{parmar2014survey}. This single-operation behavior of the aforementioned schemes is known as partial homomorphic encryption (PHE) and has limited their applications in practical scenarios \cite{reddy2023design}.  To overcome the limitations of PHE, Craig Gentry introduced the first fully homomorphic encryption (FHE) scheme \cite{van2010fully},  which allowed both addition and multiplication operations on encrypted messages. For example, it enabled expressions like:
\begin{equation}
\text{Enc}(m_1), \text{Enc}(m_2) = \text{Enc}(m_1^2 + m_2 \cdot m_1)
\end{equation}

Gentry’s FHE scheme is based on lattice-based cryptography and bootstrapping procedures. Lattice-based cryptography ensures data privacy by using complex mathematical problems such as Learning with Errors (LWE) whereas Bootstrapping is a procedure that refreshes ciphertexts (encrypted messages) to minimize the noise growth during encryption.

\begin{table*}[h]
\centering
\caption{Focused Comparison of BGV and CKKS}
\label{tab:Table1}
\begin{tabular}{@{}lll@{}}
\toprule
\textbf{Aspect}         & \textbf{BGV}                                     & \textbf{CKKS}                                      \\ \midrule
Primary Use Case        & Integer values         & Floating-point values  \\ 
Arithmetic Type         & Modular arithmetics                & Approximate arithmetics                        \\ 
Key Features            & Batching technique                  & Scaling factor                  \\ 
Applications            & Financial systems, cryptographic protocols.      & Machine learning, data analytics.                \\ \bottomrule
\end{tabular}
\end{table*}

\subsection{FHE parameters}
The adoption and effectiveness of Fully Homomorphic Encryption (FHE) in cryptographic applications depend significantly on the choice and optimization of several key parameters. These parameters directly influence the performance, security, and practicality of FHE schemes. Crucial parameters include the polymodulus degree, ciphertext modulus, plaintext modulus, scaling factor, and operation depth, all of which must be carefully chosen to balance security requirements and computational efficiency in FHE implementations \cite{bossuat2024security}, \cite{albrecht2021homomorphic}.

 Polymodulus Degree. The polymodulus degree, denoted by \( n \), determines the size of the polynomial ring used in the encryption process. It is critical in defining the complexity and security of the encryption scheme. The choice of \( n \) impacts both the computational overhead and the level of security against attacks. In ring theory, a polynomial ring R is constructed using polynomials \(f(X)\), where the coefficients belong to certain rings, and is defined as:

\begin{equation}
R = \mathbb{Z}[X]/(X^n + 1)
\end{equation}

The degree \( n \) determines the number of coefficients involved in the polynomial, thus influencing the size of the encryption parameters and the computational resources required for encryption and decryption.

Ciphertext Modulus. The ciphertext modulus  \( q \) is a key parameter that influences the noise growth during homomorphic operations. Larger ciphertext moduli enable greater noise tolerance, allowing more operations to be performed on encrypted data before the ciphertext becomes corrupted.

Plaintext Modulus. The plaintext modulus  \( t \) determines the size of the values that can be encrypted, directly impacting the range of data that can be securely handled by the encryption scheme.

Scaling Factor. The scaling factor \( \Delta \) plays a vital role in controlling the noise growth during homomorphic operations, affecting the number of operations that can be performed on the ciphertext before it becomes too noisy for accurate decryption.

Operation Depth. The depth of operations \( L \) refers to the number of homomorphic operations that can be performed on the ciphertext before the noise grows to an irreparable level. This parameter is crucial in determining how many layers of computation are feasible within the encryption framework. These parameters—\( n \), \( q \), \( t \), \( \Delta \), and \( L \)—must be carefully optimized to balance performance and security in FHE schemes.

\subsection{FHE schemes: BGV and CKKS}
Following Gentry’s foundational approach, several Fully Homomorphic Encryption (FHE) schemes have been developed, with BGV and CKKS being two of the most widely used for practical applications\cite{kim2023general}. The Brakerski-Gentry-Vaikuntanathan (BGV) scheme, introduced in 2012, is designed for efficient arithmetic operations, making it particularly suitable for exact computations, such as in cryptographic applications \cite{putra2023strix}.
 In 2017, the Cheon-Kim-Kim-Song (CKKS) scheme was introduced, which supports homomorphic computations on real numbers. CKKS employs approximation arithmetic, converting floating-point values into large integers to facilitate efficient homomorphic operations.  \cite{sathishkumar2024improving}, \cite{pan2024fedshe}. A key feature of CKKS is the scaling factor which ensures the precision of approximate calculations. This makes CKKS particularly well-suited for applications involving machine learning and other approximate computations, where exact precision is not always required.
Table \ref{tab:Table1} highlights the key differences between the BGV and CKKS schemes.

\begin{figure*}[ht]
    \centering
    % First figure
    \begin{minipage}{0.45\textwidth}
        \centering
        \includegraphics[width=\textwidth, height=9cm]{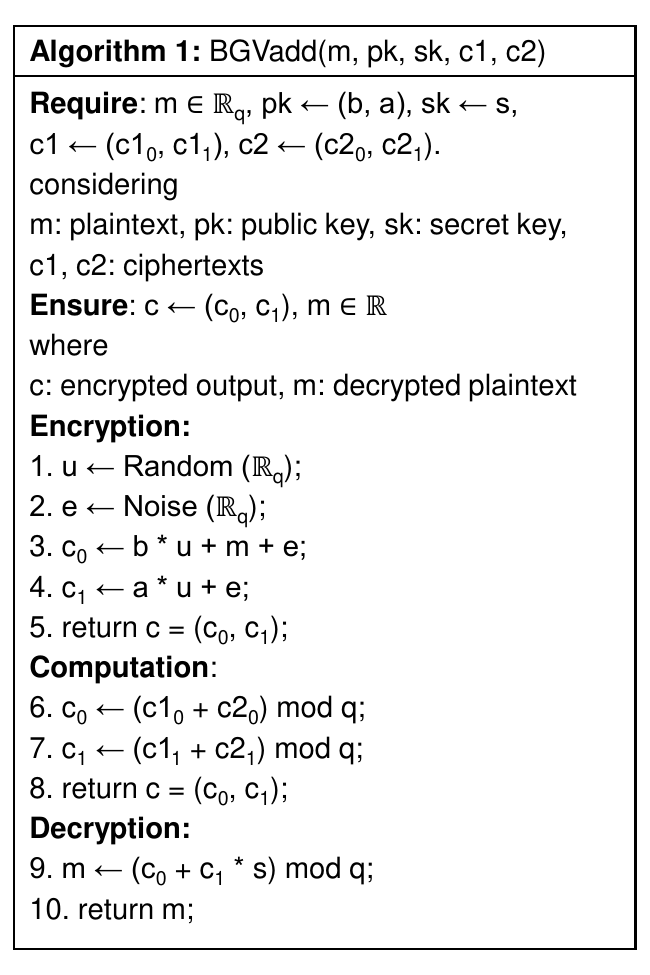} % Replace with your first image file
        
    \end{minipage}
    \hfill % Space between figures
    % Second figure
    \begin{minipage}{0.45\textwidth}
        \centering
        \includegraphics[width=\textwidth, height=9cm]{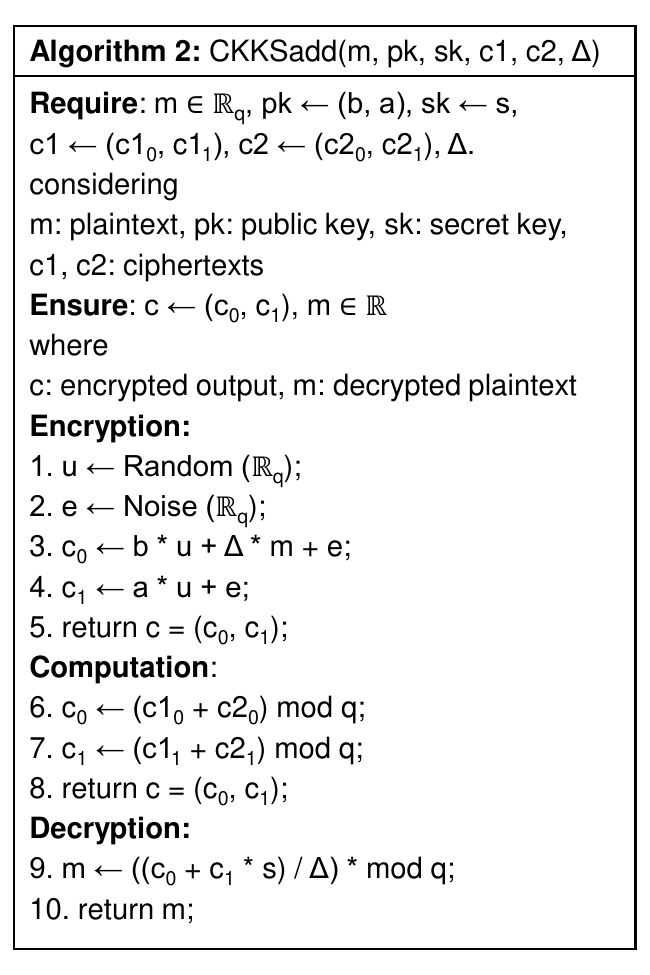}
    \end{minipage}
\end{figure*}

\subsection{FHE libraries: Microsoft SEAL and OpenFHE}
To enable the practical deployment of Fully Homomorphic Encryption (FHE) schemes, several libraries have been introduced, with Microsoft SEAL and OpenFHE being among the most widely used.
 Microsoft SEAL is developed by Microsoft Research and is known for its user-friendly features and predefined FHE parameters, making it accessible to both researchers and practitioners \cite{valera2024empirical}.The library incorporates advanced techniques like memory pooling and multi-threading to optimize resource utilization and performance. SEAL also supports cross-platform deployment, ensuring interoperability across various operating systems, including Windows, Linux, and macOS, which enhances its versatility in real-world applications. OpenFHE, on the other hand, is a research-driven library that also supports FHE implementations across multiple platforms. \cite{al2022OpenFHE}. 
 While both libraries differ in their design and architecture, they share a common goal: to facilitate the practical implementation of FHE schemes in real-world applications. Microsoft SEAL is well-suited for straightforward use cases, while OpenFHE provides more flexibility for cutting-edge research and complex application designs. This comparison provided valuable insights into the efficiency of these libraries under different conditions.

\section{Related Work}
Different studies have analyzed the performance of FHE libraries across different FHE schemes. The authors in \cite{aguilar2019comparison} compared different FHE libraries such as HElib, SEAL, and FV-NFLlib and analyzed their performance with large plaintext moduli. Similarly, Jiang et. al evaluated these libraries under various FHE operations and provided a detailed analysis of their computational performance \cite{jiang2022fhebench}. In another study, Gouert et al. examined the performance and usability of various FHE libraries, offering a comprehensive overview of their strengths and limitations \cite{gouert2023sok}.While these studies provide valuable insights into the computational performance of FHE libraries, they do not address critical aspects such as memory usage and scalability across different platforms. These factors are essential for understanding the overall efficiency and adaptability of FHE libraries in diverse real-world environments.
Recently, various research studies have also discussed the performance of specific FHE schemes to better understand FHE libraries. Chillotti et al. analysed the TFHE scheme to improve the performance of homomorphic operations \cite{iliashenko2021faster}. The authors in \cite{clet2021bfv} reviewed different FHE schemes such as BFV, CKKS, and TFHE schemes, and evaluated neural networks based on requirements. Ma et al. compared PyTFHE which is a Python wrapper of TFHE scheme, with other frameworks to highlight improvements in HE computations \cite{ma2023pytfhe}. However, these studies did not explore the cross-platform performance of FHE schemes, especially in edge-cutting FHE libraries such as OpenFHE and SEAL. 
Several studies have also compared the performance of FHE libraries across various applications. Suzuki et al. analyzed the performance of SEAL and OpenFHE for convolutional neural networks, focusing on the trade-offs between processing latency and model accuracy \cite{zhu2023performance}. Moreover, Zhai et al. presented GPU acceleration for encrypted computation using SEAL, focusing on hardware-specific optimizations \cite{zhai2022accelerating}. While these studies provide valuable benchmarks for different FHE libraries, they fall short in addressing the cross-platform performance, memory usage, and operational depth of these libraries.  The present study seeks to address these gaps by analyzing the SEAL and OpenFHE libraries across multiple platforms, focusing on both computation time and memory consumption. By providing new insights into these critical aspects, our work contributes to a deeper understanding of the performance of FHE libraries and promotes more efficient solutions for real-world FHE applications.

\section{Proposed Methodology}
This section outlines a benchmarking methodology designed to compare the performance of two FHE libraries: SEAL and OpenFHE. The study evaluates two key performance metrics—execution time and memory usage—across two operating systems, Windows and Linux. The analysis focuses on two widely adopted FHE schemes, BGV and CKKS, ensuring a comprehensive comparison. All experiments were conducted on a a laptop featuring an Intel Core i7-8550U processor, 16 GB of RAM, and a 512 GB SSD. To ensure uniformity and reliability of results, identical parameter settings were applied across all platforms. Performance evaluation was based on two fundamental FHE operations: addition and multiplication. The experiments explored various configurations parameters, including polymodulus degree, coefficient modulus, plaintext modulus, and scaling factor.
The evaluation of the addition operations was conducted using a fixed set of operation depths (50, 100, 150, 200, 250, and 300) across polymodulus degrees of 13, 14, and 15. In contrast, the multiplication operations required dynamic parameter configurations due to their significantly higher computational complexity. As the operation depth increased, the polymodulus degree was scaled accordingly to maintain computational accuracy and ensure correctness. Table \ref{tab:Table2} illustrates the correlation between operation depth and polymodulus degree, highlighting the adjustments necessary to accommodate deeper operations. Moreover, the coefficient modulus chain and scaling factor were also updated to accommodate the increasing computational complexity that resulted from the approximation arithmetic technique inherent in the CKKS scheme. 

\begin{table}[ht]
\centering
\caption{Correlation between operation depth and polymodulus degree.}
\label{tab:Table2}
\begin{tabular}{p{2cm}cccc}
\toprule
\textbf{Operation Depth (L)} & \multicolumn{4}{c}{\textbf{ Polymodulus Degree (q)}} \\ 
\cmidrule(r){2-5}
                         & \multicolumn{2}{c}{\textbf{SEAL}} & \multicolumn{2}{c}{\textbf{OpenFHE}} \\ 
\cmidrule(r){2-3} \cmidrule(l){4-5}
                         & \textbf{BGV} & \textbf{CKKS}      & \textbf{BGV}         & \textbf{CKKS}         \\ 
\midrule
2                      & 12           & 13                 & 14                   & 14                   \\ 
3                     & 13           & 13                 & 14                   & 14                   \\ 
4                        & 13           & 14                 & 14                   & 14                   \\ 
5                        & 13           & 14                 & 14                   & 15                   \\ 
6                        & 13           & 14                 & 14                   & 15                   \\ 
7                        & 14           & 14                 & 14                   & 15                   \\ 
8                        & 14           & 14                 & 14                   & 15                   \\ 
9                        & 14           & 15                 & 14                   & 15                   \\ 
10                       & 14           & 15                 & 14                   & 15                   \\ 
11                       & 14           & 15                 & 14                   & 16                   \\ 
12                       & 14           & 15                 & 14                   & 16                   \\ 
13                       & 15           & 15                 & 15                   & 16                   \\ 
14                       & 15           & 15                 & 15                   & 16                   \\ 
15                       & 15           & 15                 & 15                   & 16                   \\ 
16                       & 15           & 15                 & 15                   & 16                   \\ 
17                       & 15           & 15                 & 15                   & 16                   \\ 
18                       & 15           & 15                 & 15                   & 16                   \\ 
19                       & 15           & 15                 & 15                   & 16                   \\ 
20                       & 15           & 15                 & 15                   & 16                   \\ 
\bottomrule
\end{tabular}

\end{table}

\section{Results and Discussion}
 This section presents a detailed benchmarking analysis of the SEAL and OpenFHE libraries, focusing on their performance under the BGV and CKKS schemes. The discussion is structured into two subsections, each dedicated to evaluating the performance of  core homomorphic operations: FHE addition and FHE multiplication. 

\subsection{FHE-Addition Performance}
\begin{figure*}[ht]
    \centering
    % First figure
    \begin{minipage}{0.45\textwidth}
        \centering
        \includegraphics[width=\textwidth]{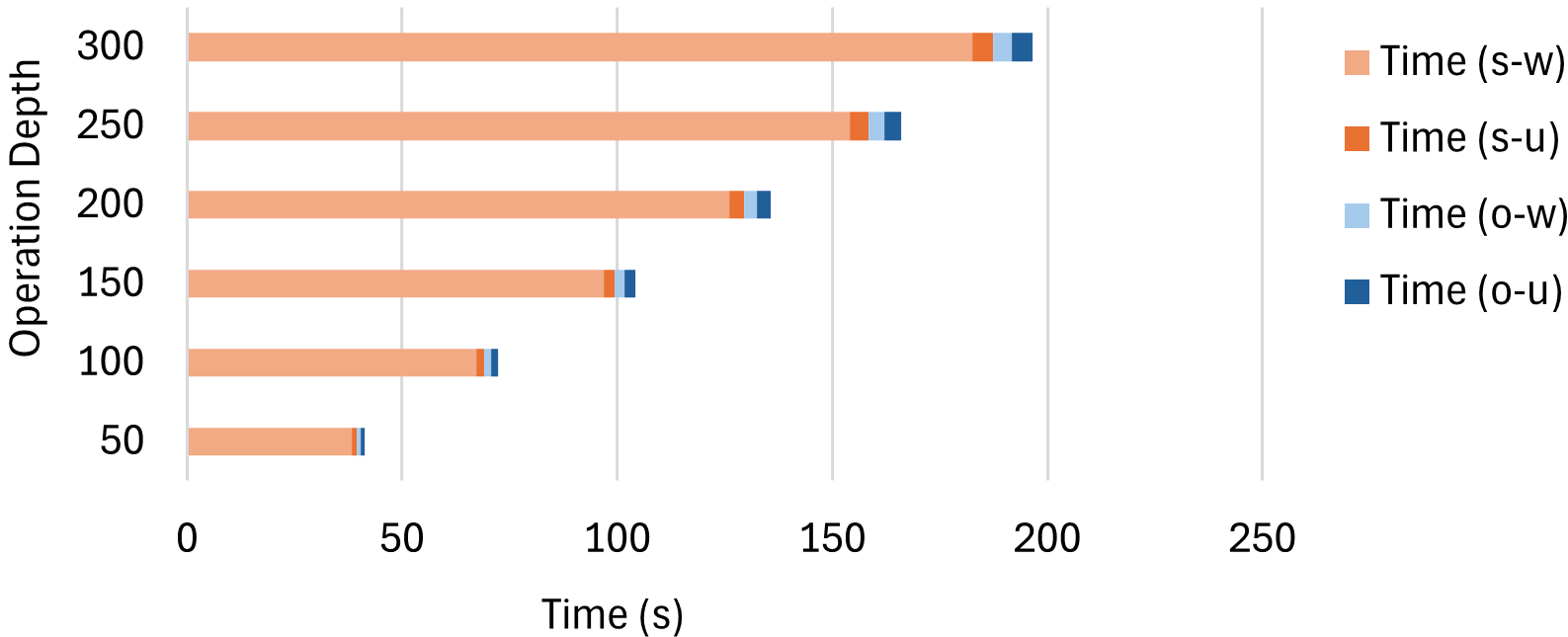} % Replace with your first image file
        \caption{Execution Time for FHE-addition in BGV.}
        \label{fig:fig4}
    \end{minipage}
    \hfill % Space between figures
    % Second figure
    \begin{minipage}{0.45\textwidth}
        \centering
        \includegraphics[width=\textwidth]{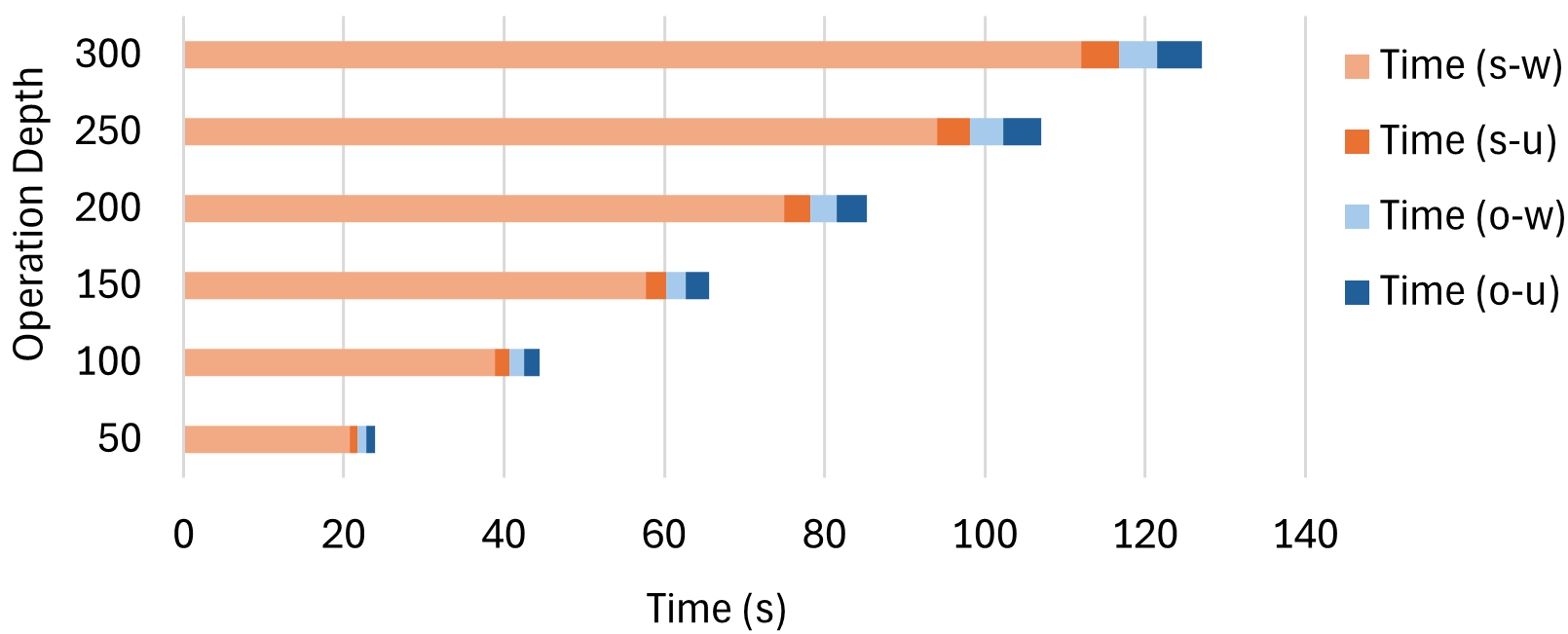} % Replace with your second image file
        \caption{Execution Time for FHE-addition in CKKS.}
        \label{fig:fig5}
    \end{minipage}
\end{figure*}

\begin{figure*}[ht]
    \centering
    % First figure
    \begin{minipage}{0.45\textwidth}
        \centering
        \includegraphics[width=\textwidth]{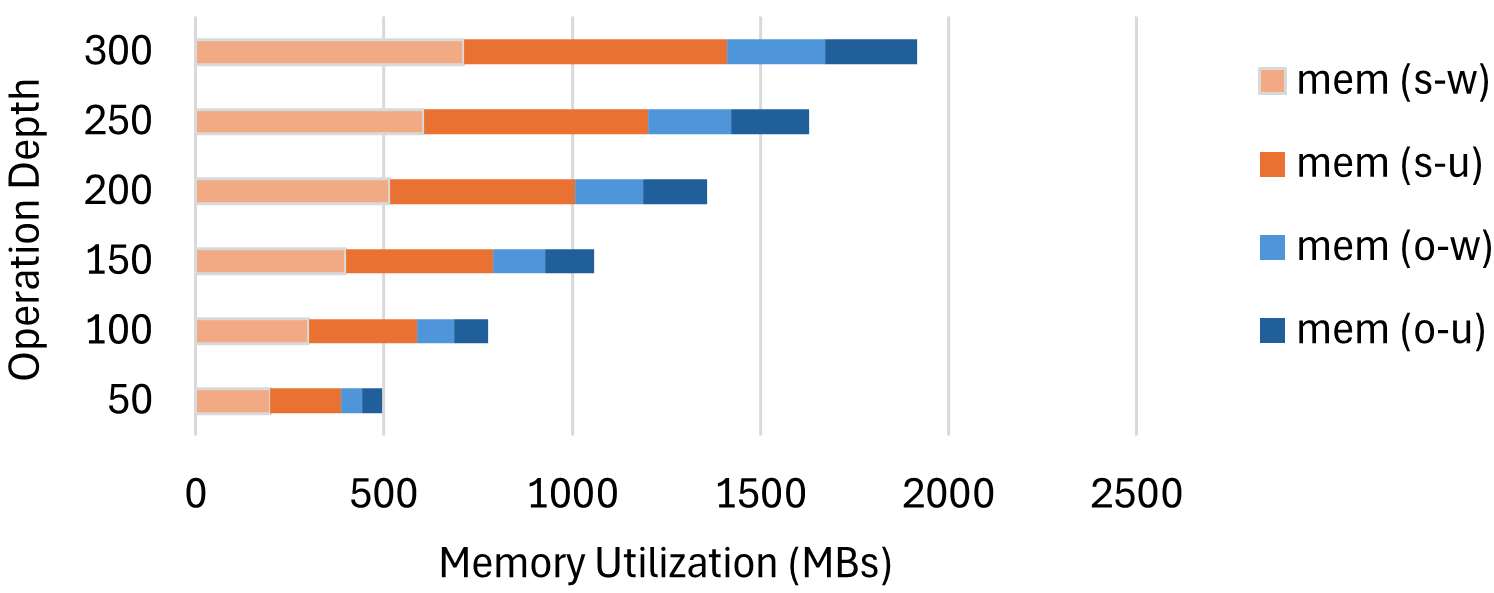} % Replace with your first image file
        \caption{Memory Utilization for FHE-addition in BGV.}
        \label{fig:fig6}
    \end{minipage}
    \hfill % Space between figures
    % Second figure
    \begin{minipage}{0.45\textwidth}
        \centering
        \includegraphics[width=\textwidth]{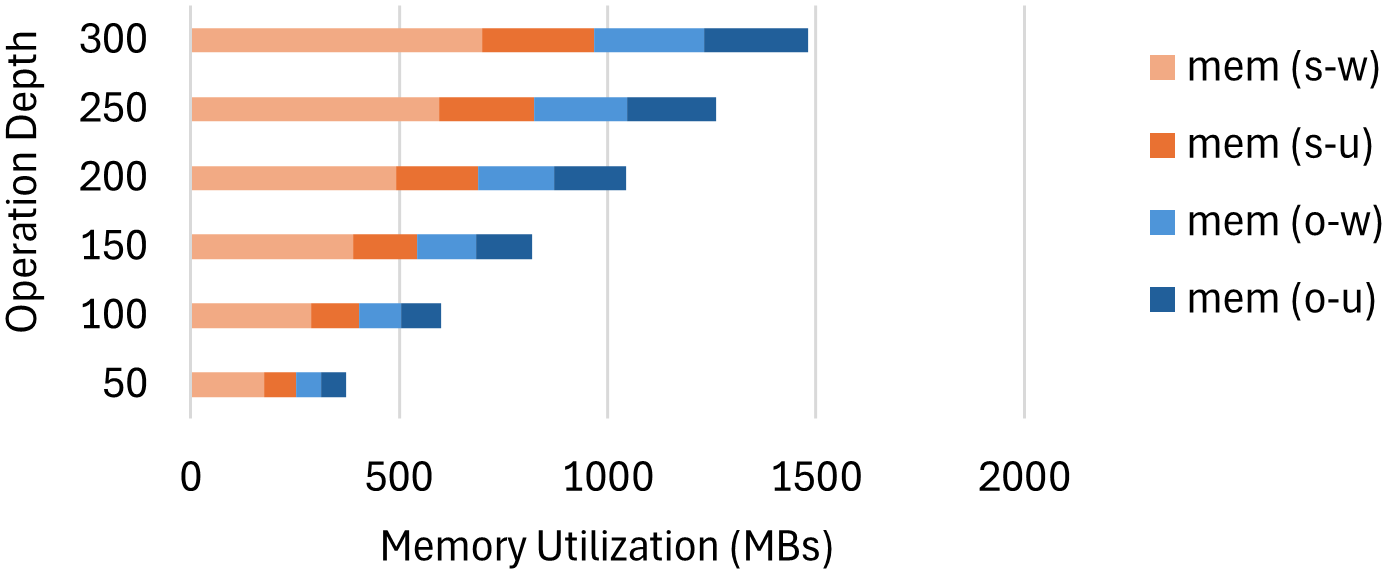} % Replace with your second image file
        \caption{Memory Utilization for FHE-addition in CKKS.}
        \label{fig:fig7}
    \end{minipage}
\end{figure*}

 This section presents a detailed benchmarking analysis of the SEAL and OpenFHE libraries, focusing on their performance under the BGV and CKKS schemes. The discussion is structured into two subsections, each dedicated to evaluating the performance of  core homomorphic operations: FHE addition and FHE multiplication, outlined in Figures \ref{fig:fig4} to \ref{fig:fig7}.

Figure \ref{fig:fig4} shows the execution time of FHE addition under the BGV scheme as a function of operation depth. Execution time increased gradually with operation depth for both libraries. SEAL initially performed well but exhibited a sharp rise in execution time on Windows as the operation depth grew. At a depth of 300, SEAL's execution time on Windows reached 182 seconds, highlighting significant scalability challenges. In contrast, OpenFHE demonstrated consistently faster execution times on both Windows and Linux, with Linux outperforming Windows in all cases, further emphasizing OpenFHE's optimized performance. Similarly, the execution times in CKKS were consistently higher than in BGV due to the further computational complexity introduced by the approximation technique. However, OpenFHE on Linux still has the additional benefit of completing 300 FHE additions in almost 4 seconds, compared to SEAL’s 7 seconds (approx.) on the same platform. Figure \ref{fig:fig5} presents the execution time of the FHE-addition operation in CKKS across all configurations.

Moreover, \ref{fig:fig6} outlines a similar pattern for memory usage. In the BGV, SEAL performed slightly better on Linux compared to Windows. For instance, the memory usage at depth 300 was 690 MB on Linux and 715 MB on Windows. In contrast, OpenFHE demonstrated considerably better memory usage across all configurations with 240 MB utilized for the same operation depth. Similarly, the CKKS consumed more memory than BGV for both libraries across all platforms, as shown in \ref{fig:fig7}. SEAL demanded the highest memory on Windows, whereas OpenFHE remained memory efficient even at larger operation depths. This highlights its potential to effectively manage resource demands for CKKS.

\subsection{FHE-Multiplication Performance}
\begin{figure*}[ht]
    \centering
    % First figure
    \begin{minipage}{0.45\textwidth}
        \centering
        \includegraphics[width=\textwidth]{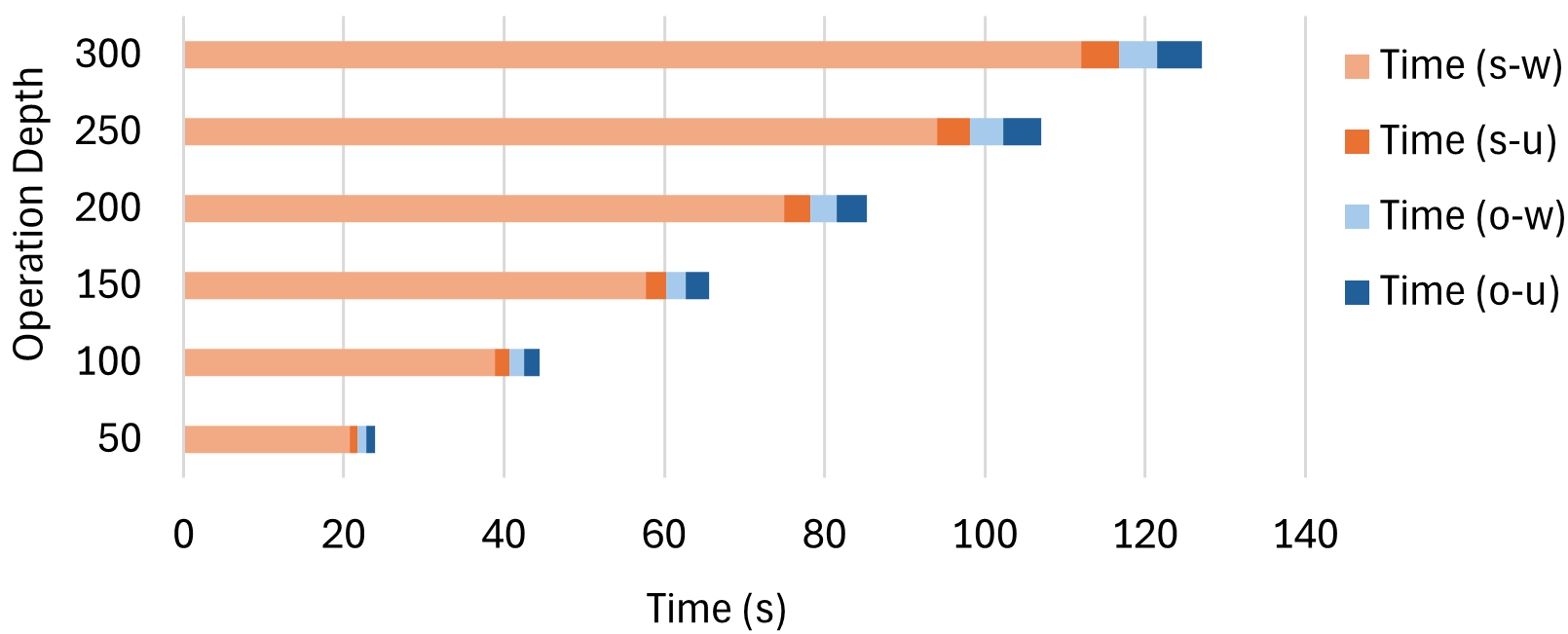} % Replace with your first image file
        \caption{Execution Time for FHE-multiplication in BGV.}
        \label{fig:fig8}
    \end{minipage}
    \hfill % Space between figures
    % Second figure
    \begin{minipage}{0.45\textwidth}
        \centering
        \includegraphics[width=\textwidth]{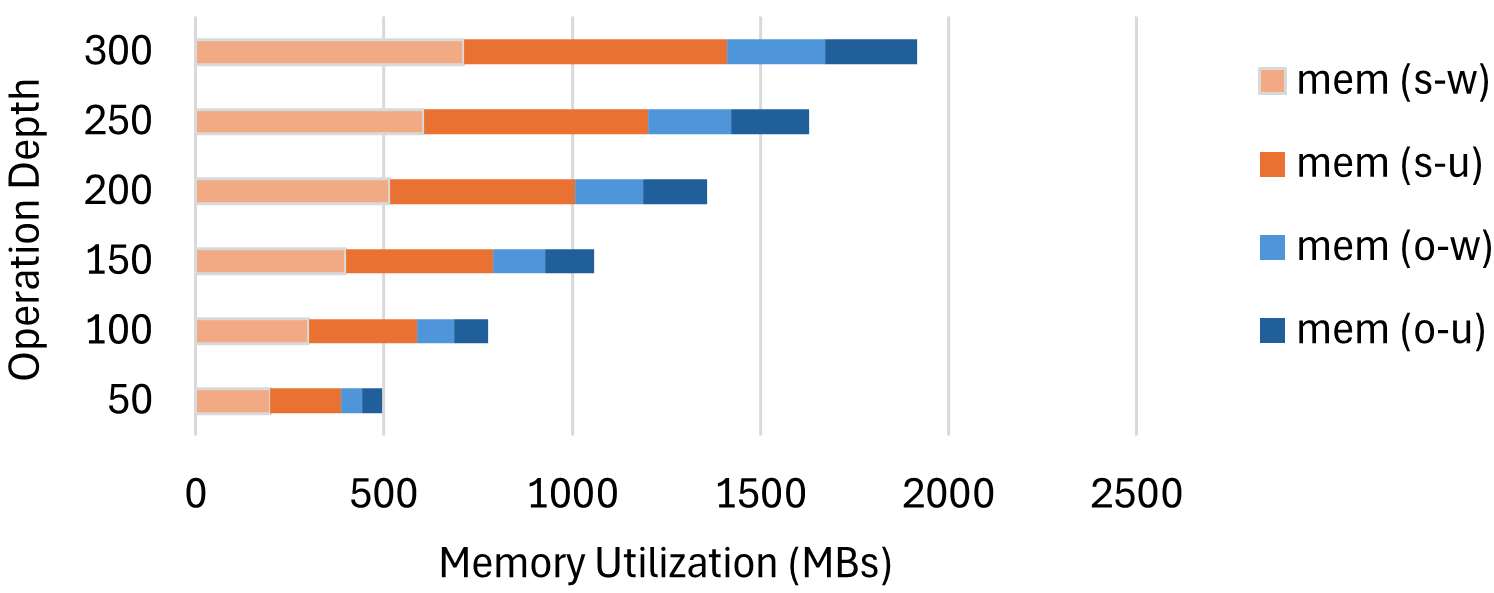} % Replace with your second image file
        \caption{Execution Time for FHE-multiplication in CKKS.}
        \label{fig:fig9}
    \end{minipage}
\end{figure*}

\begin{figure*}[ht]
    \centering
    % First figure
    \begin{minipage}{0.45\textwidth}
        \centering
        \includegraphics[width=\textwidth]{Fig10.png} % Replace with your first image file
        \caption{Memory Utilization for FHE-multiplication in BGV.}
        \label{fig:fig10}
    \end{minipage}
    \hfill % Space between figures
    % Second figure
    \begin{minipage}{0.45\textwidth}
        \centering
        \includegraphics[width=\textwidth]{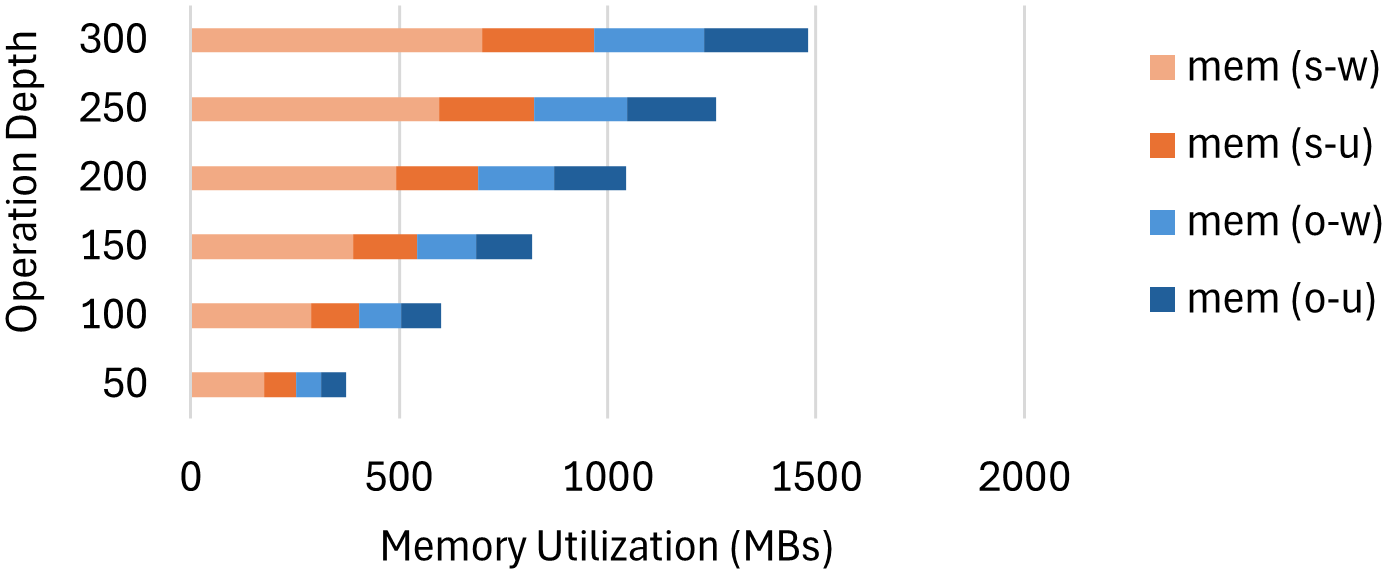} % Replace with your second image file
        \caption{Memory Utilization for FHE-multiplication in CKKS.}
        \label{fig:fig11}
    \end{minipage}
\end{figure*}

The FHE-multiplication analysis offers a comprehensive overview of the computational efficiency and memory usage of the aforementioned FHE libraries shown in Figures \ref{fig:fig8} to \ref{fig:fig11}.
The execution time for multiplication operations in BGV has risen continuously with operation depth across both libraries. The details are outlined in Figure \ref{fig:fig8}. Overall, SEAL took significantly longer execution times on Windows. For instance, at operation depth 20, SEAL took more than 530 seconds to execute on Windows, which indicates the future challenges for scalability. Meanwhile, OpenFHE outperformed across each configuration, executing 20 multiplications in approximately 7 seconds on Linux, which significantly overtook SEAL on both platforms. Similarly, the execution time for CKKS was generally higher than in BG as described in FHE-addition. The details are shown in Figure \ref{fig:fig9}. SEAL on Windows took about 300 seconds for depth 20, whereas OpenFHE on Linux exhibited the same operations in 12 seconds, which shows a clear performance efficiency.

Moreover, the memory usage in BGV and CKKS is highlighted in Figures 7 and 8, respectively. It is observed that the memory demand for SEAL in BGV increased dramatically, such as for 20 FHE multiplications, which required 850 MB on Windows. However, while Linux showed slightly better performance, OpenFHE was found to be more memory efficient, requiring less than half of SEAL. For example, at depth 20, OpenFHE showcases its optimized resource management by consuming 400MB on the Linux platform, as shown in Figure \ref{fig:fig10}. Similarly, for the memory usage in CKKS, SEAL exceeded 1200 MB at depth 20 on Windows, making it the most resource-intensive configuration, as illustrated in Figure \ref{fig:fig11}. Whereas, OpenFHE has again proven its efficiency by requiring under 800 MB on Linux for the same operation depth.

Furthermore, a significant increase in execution time and memory usage was observed with each transition to a higher polymodulus degree $(n)$ to manage higher depths of multiplication operations. For each transition, SEAL showed more resource growth while OpenFHE demonstrated fewer fluctuations. For instance, in BGV, at operation depth 9, for transitioning from polymodulus degree $2^{13}$ to $2^{14}$, SEAL on Windows consumed $5x$ more execution time than the regular pattern. Whereas, at depth 13, transitioning from polymodulus degree $2^{14}$ to $2^{15}$ in OpenFHE resulted in only a $2x$ increase on Linux, which confirms its improved scalability.

\subsection{Comparative Insights}
Overall, OpenFHE continuously outperformed SEAL in terms of execution time and memory usage. This outperforming behavior of OpenFHE became more evident at higher operation depths, particularly in CKKS to handle the scaling factors in managing computational overhead. However, the higher computational demand of SEAL may limit its scalability in advanced applications. The performance of SEAL was also greatly affected by the platform choice such as Linux consistently performed better than Windows.  Hence, OpenFHE is an optimal choice for improved resource utilization which makes it suitable for cryptographic applications.

\section{Conclusion}
Our study benchmarks two widely used FHE libraries, SEAL and OpenFHE under BGV and CKKS schemes. This benchmarking provides new insights into computation overhead and the scalability of these libraries in a cross-platform approach. Our findings reveal that OpenFHE consistently outperformed SEAL in all configurations, where Linux appeared to be a more efficient platform for each operation setting. In the future, this study can be extended to include more libraries using different FHE schemes that would help researchers gain comprehensive insights into FHE advancements. Moreover, different hardware settings, such as GPUs or specialized accelerators, can be involved to highlight the performance of homomorphic operations. Furthermore, the evaluation of multiple FHE schemes in neural networks using different libraries would offer new research directions in implementing privacy-preserving applications. 

\bibliographystyle{IEEEtran}
\bibliography{v3}

\end{document}